# Temporal Resolution of Measurements and the Effects on Calibrating Building Energy Models

Fazel Khayatian*, Andrew Bollinger, Philipp Heer
Urban Energy Systems Laboratory, Swiss Federal Laboratories for Materials Science and Technology, Empa, Dübendorf, Switzerland
*fazel.khayatian@empa.ch

## Abstract
With the recent interest in installing building energy management systems, the availability of data enables calibration of building energy models. This study compares calibration on eight different temporal resolutions and contrasts the benefits and drawbacks of each. While heating and cooling calibration show less sensitivity to temporal resolutions, the performance of calibrating electricity energy consumption and domestic hot water greatly varies depending on the temporal resolution.

## Introduction

Building simulation software are extensively used for estimating the energy demand of buildings. The information fed into building energy models are often simplified or generalized, given their roots in standards, regulations, or the literature. Whilst simulations with conventional assumptions could potentially provide useful information for parametric and comparative studies, drawing conclusions on the performance of a system is highly dependent on the validity of the prior assumptions. Therefore, calibration of building energy models has become an integral part of the building energy simulation pipeline, conditioned to the availability of measurements (Asadi, Mostavi, Boussaa, & Indaganti, 2019).

A calibrated building energy model, not only creates a reliable baseline to assess modifications to an existing building (such as retrofit) (Heo, Choudhary, & Augenbroe, 2012), but also enriches the prior assumptions that that were initially utilized to set up the energy model (Chong, Xu, Chao, & Ngo, 2019). Regardless of the drivers, availability of metered usage is the prerequisite for calibrating a building energy model. Whilst case studies have reported metered energy use in monthly intervals (Garrett, New, & Chandler, 2013) (AlAjmi, Abou-Ziyan, & Ghoneim, 2016) (Williams & Gomez, 2016), recent studies report a growing interest toward hourly or sub-hourly measurements (Reddy, 2006) (Raftery, Keane, & O'Donnell, 2011) (Coakley, Raftery, & Keane, 2014). Opting for fine temporal resolution in measurements is propelled by the introduction of Building Energy Management Systems (BEMS), as well as Smart Grids (SG) (Hurtado, Nguyen, & Kling, 2014) (Hurtado, Nguyen, & Kling, 2015). Whilst BEMS typically suffice to 30 or 15 minutes data logging (Ahmad, Mourshed, Mundow, Sisinni, & Rezgui, 2016), it is argued that one minute data sampling could be beneficial for a deep study into the electrical loads (Lanzisera et al., 2013). It is important to note that BEMS and SG metering were not originally aimed at model calibration problems. Studies highlight that resorting to coarse temporal resolutions for calibration can greatly affect the performance of the tuned model, specifically when benchmarked against sub-hourly measurements (Kristensen, Choudhary, & Petersen, 2017). On the other hand, it is argued that even hourly profiles could be too detailed (and perhaps noisy) for comparisons (Chaudhary et al., 2016).

Therefore, a systematic study on utilizing BEMS data for calibration purposes is necessary. As an original contribution, this paper contrasts the benefits and drawback of calibrating a building energy model with coarse and fine temporal resolutions.

## Methods

Given that the thresholds for calibration can be difficult to meet, opting for a one-off Latin Hypercube Sampling (LHS) cannot guarantee the intended accuracy, unless an excessive number of simulations is executed. Conditional subset simulation with adaptive threshold can overcome this challenge (Bect, Li, & Vazquez, 2017). Subset simulation is a special case of the Metropolis-Hastings algorithm that yields a distribution of interest from a set of weakly-informative priors (Papaioannou, Betz, Zwirglmaier, & Straub, 2015). The procedure was originally designed to reach low probabilities in reliability studies, but is also proven suitable for calibration purposes (Meshkinkiya, Loonen, Paolini, & Hensen, 2019). Subset simulation treats the model as a black-box, and therefore, only the inputs (random variables) and the criteria (thresholds) are required to explore the uncertain input space. However, unlike black-box surrogate models that fit a separate estimator onto inputs and responses, subset simulation infers a prior space from the model's responses, and iteratively narrows the prior space through sampling. To overcome the computational burden of sampling, subset simulation resorts to an efficient sampling method based on the Markov chain process.

## Sampling from mixture models

Performing a coarse grid search with Monte Carlo simulation has proved adequately accurate for calibration purposes, regardless of the measurements' temporal resolution. In contrast, the precision of surrogate-based calibration can greatly vary depending on the data's temporal resolution. Resorting to pseudorandom simulation eliminates the need for a surrogate, including the epistemic uncertainty that is stemmed from the surrogate's imprecision. Monte Carlo sampling is suitable for inferring multidimensional prior spaces from multimodal responses, i.e., in this study building energy model calibration based on multiple measurements. This could be useful for simultaneous tuning of heating, cooling, and indoor air temperature, given that all are responses to a mixture of occupancy rate, lighting, and equipment power.

## Efficient sampling

As mentioned before, performing calibration with Monte Carlo sampling can have advantages over using surrogate models or emulators. However, this comes at the cost of additional computation time and resources. To overcome this challenge, subset simulation resorts to Markov Chain Monte Carlo, in which the properties of a prior distribution is incrementally inferred by iterative sampling from the intended posterior space. The high efficiency of subset simulation is particularly appealing for calibration of complex energy models, as well as high temporal resolutions which suffer from lengthy computations.

*Table 1: Calibration algorithm pseudocode*

```
1:  procedure SUBSET-SIMULATION
2:     ℝ := plausible ranges for all n variables
3:     υ := generate m random samples with LHS
4:     simulate all members of υ
5:     calculate CVRMSE and NMBE for all members of υ
6:     θ₁ := min(CVRMSE)
7:     θ₂ := min(NMBE)
8:     While θ₁>Threshold1 & θ₂>Threshold2 do
9:        η₁ := |CVRMSE – Threshold1|
10:       η₂ := |NMBE – Threshold2|
11:       Rescale η₁ and η₂ between 0 and 1
12:       ή := √(η₁² + η₂²)
13:       sort ή from small to large
14:       sort υ based on ή
15:       ύ := υ(1:k)
16:       ϕ := fit a mixture model on ύ
17:       truncate model ϕ with ranges of ℝ
18:       υ := generate m new samples from ϕ
19:       simulate all members of υ
20:       calculate CVRMSE and NMBE for all members of υ
21:       θ₁ := min(CVRMSE)
22:       θ₂ := min(NMBE)
23:    return ύ
24: end procedure
```

## Calibration measures

The Coefficient of Variation of the Root Mean Squared Error (CVRMSE) and the Normalized Mean Bias Error (NMBE) are proposed by ASHRAE as the conventional criteria for calibrating the model (Haberl, Claridge, & Culp, 2005), and therefore, are adopted to evaluate the calibration processes in this study. Subset simulation iteratively minimizes the distance between the perceived performance of the simulations, and the intended measure for performance (Table 1). Therefore, the loop continues to resample from the prior space, until at least one sample exceeds all performance thresholds in the posterior space. For instance, if CVRMSE of 30% and NMBE of 10% are the intended thresholds, the loop will stop iterating as soon as a single sample returns both CVRMSE≤0.3 and NMBE≤0.1 simultaneously.

Whilst regulations have introduced certain threshold values to define a "*calibrated energy model*", using a fixed threshold value is ineffective and misleading to evaluate information loss. Therefore, a new measure is required to set the stopping criteria of the subset simulation loop, one that is adaptable to each temporal resolution. In other words, the stopping criteria should guarantee that the energy model of each temporal resolution is sufficiently tuned. To address this challenge, this study monitors the improvement in the calibration progress. This will allow us to sufficiently calibrate each energy model, to the point from which improvements are considered trivial. Namely, the calibration loop will stop if the improvement in both metrics is smaller than 1%. This will guarantee that each calibration procedure has converged, and sets a unified evaluation criteria for all temporal resolutions.

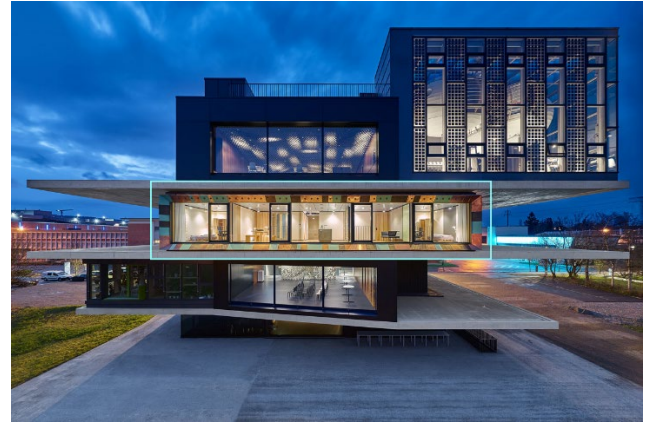

*Figure 1: Elevation of the case study residential unit. © Zooey Braun, Stuttgart.*

## Case study

The case study (Figure 1) is a residential unit of the NEST demonstrator at Empa, in Switzerland (Richner, Heer, Largo, Marchesi, & Zimmermann, 2017). The unit is connected to a central distribution network, from which long-term metering of heating, cooling, electricity and

Domestic Hot Water (DHW) is available for each unit. An energy model of the unit is created using DesignBuilder version 4.7, and the EnegryPlus input file is extracted for calibration. The unit consists of two bedrooms, two bathrooms (merged in the modelling), an entry hall, as well as an open plan living space including a kitchen, among which the entry hall and the storage space are not conditioned (Figure 2). Modifications to the EnergyPlus model for calibration are applied through the Python eppy library (Santosh Philip, 2016).

calibration. Since consumption is metered at the connection point between the unit and the central energy distribution network, heating and cooling are set to ideal loads and simulated as a district heating and cooling system for the entire unit.

**Data processing**

Data is collected in 1-minute intervals for heating, cooling, electricity as well as DHW. Weather data is collected onsite including dry-bulb and dew-point air temperatures, relative humidity, absolute pressure, wind speed and direction, as well as global horizontal solar radiation. Direct and diffuse radiation are derived from global horizontal solar radiation using the Reindl method (Reindl, Beckman, & Duffie, 1990). Missing values up to 3 hours in energy consumption are resampled with linear interpolation. Larger gaps in the energy consumption are left as missing values and excluded from CVRMSE and NMBE calculations. Missing weather data are infilled from a nearby weather station provided by the Swiss Federal Office of Meteorology and Climatology (MeteoSwiss, 2020). Five separate EnergyPlus Weather Files (epw) are generated for minutely to hourly simulations. Daily, 6-hour and monthly simulations are performed with hourly weather data.

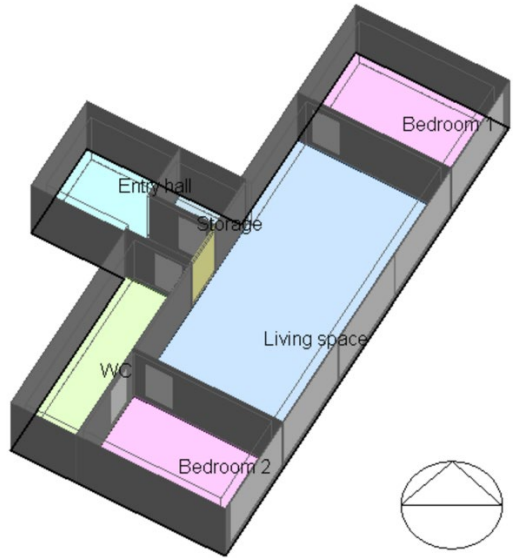

*Figure 2: Energy model of the case study residential unit.*

*Table 2: Temporal resolution of calibration*

| Calibration temporal resolution | Simulation time-step | Measurement interval |
|---|---|---|
| 1-minute | 1-minute | 1-minute |
| 5-minutes | 5-minutes | Aggregated from 1-minute measurements |
| 15-minutes | 15-minutes | |
| 30-minutes | 30-minutes | |
| Hourly | Hourly | |
| 6-Hours | Aggregated from Hourly simulations | |
| Daily | | |
| Monthly | | |

Eight different calibrations are performed with various time-steps as described in Table 2. The temporal resolutions are chosen from a number of studies which deal with different sampling times for various applications, including load disaggregation (5-minutes) (Perez et al., 2014), predictive control (15-minutes) (Sturzenegger, Gyalistras, Morari, & Smith, 2015), building energy saving (30minutes) (Ferreira & Fleming, 2014), as well as calibration (hourly, 6-hours, daily, monthly) (Fabrizio & Monetti, 2015). Since the maximum simulation time-step in EnergyPlus is hourly, larger intervals are obtained by resampling the hourly data. Measurements on the other hand, are collected in 1-minute intervals, and then aggregated for all other temporal resolutions of the

*Table 3: Calibrated variables and plausible ranges.*

| Variable | Unit | Range |
|---|---|---|
| Occupant heat gain density | W/m$^2$ | [0.9 , 1.1] |
| Appliance power density | W/m$^2$ | [10 , 50] |
| Lighting power density | W/m$^2$ | [2 , 5] |
| Appliance radiant fraction | % | [20 , 40] |
| Lighting radiant fraction | % | [30 , 60] |
| Ventilation rate | m$^3$/s-m$^2$ | [3.0E-4 , 9.0E-4] |
| Infiltration rate | m$^3$/s-m$^2$ | [3.0E-5 , 9.0E-5] |
| Heating set-point temperature | ºC | [18 , 24) |
| Cooling set-point temperature | ºC | [24 , 27] |
| Glass dirt correction factor | % | [0.5 , 0.9] |
| Insulation thickness (wall) | m | [0.05 , 0.01] |
| Insulation thickness (floor & ceiling) | m | [0.2 , 0.4] |
| Insulation thickness (window) | m | [5.0E-4 , 1.5E-3] |
| DHW peak flow rate | m$^3$/s | [1.0E-5 , 1.0E-4] |

**Variables**

Calibration is applied to two categories, i.e. construction and operation characteristics. Since detailed information on the unit is available, the built characteristics are assumed to be fairly accurate. Therefore thermal conductivity and solar gain through the glazed surfaces are the only construction characteristics subject to calibration. The thickness of the insulation materials are associated with uncertainty to fine-tune the thermal conductivity of the envelope. Uncertainty in the glazing transparency is reflected through the dirt correction factor of the glass (see Table 3).

Although lighting and appliances power density profiles are derived from the occupancy profile, all three are considered as separate uncertain variables. Such assumption is due to the fact that occupancy contributes to latent loads, whilst artificial lighting is controlled by illuminance sensors. Lighting and appliance radiant fractions affect the response of the heating and cooling systems to the lighting and appliance heat gain. Since electricity, heating and cooling are calibrated simultaneously, including lighting and appliance radiant fractions as uncertain variables can ensure better fine-tuning of lighting and appliance power density.

**Profiles**

Profiles of occupancy, appliance and lighting power as well as DHW, are reconstructed by mining the measurements through k-medoids clustering (Kotzur, Markewitz, Robinius, & Stolten, 2018). The Silhouette criterion is the basis for choosing the optimal number of clusters (Rousseeuw, 1987). Electricity measurements are reshaped into daily profiles, and then clustered, which form representative occupancy, lighting, and appliance profiles. The average values of each cluster is considered as the representative profile of that group, and applied to all of its member days. Except for daily and monthly measurements, a separate clustering is performed for each temporal resolution (Figure S1). Daily and monthly simulations are performed with nominal profiles as described in (Merkblatt, 2015). Similarly DHW is also clustered into daily profiles. Infiltration is assigned with a constant profile all around the year. All profiles are compiled into a comma-separated values file, and loaded into EnergyPlus as a File type profile.

As expected, moving from coarse time-steps to finer resolutions, increases the smoothness of the profiles. However, increasing the temporal resolution beyond 5-minutes can shift the time of the peak DHW between 30 and 90 minutes. Also, clustering finer temporal resolutions adds a bias to the profile's base value, effectively reducing the range between base and peak demands. The clustered 6-hour temporal resolution is the only profile which notably distorts the actual patterns.

**Results**

Simulations are executed on an Intel Xeon Gold 6244 CPU @ 3.60 GHz and parallelized in batches of 30 jobs. Details on the performance of calibrated models are provided in the Supplementary materials section (Table S1). It is observed that DHW returns low calibration errors in all temporal resolutions (Figure S2). This may be due to the fact that DHW profiles follow recurring patterns and are easily clustered into distinct groups. Moreover, DHW is zero for a considerable portion of the year. The calibration of Electricity profiles is fairly acceptable in all temporal resolutions. Cooling is also reasonably calibrated, particularly in finer temporal resolutions. However the performance of cooling (as well as heating) calibration show an unexpected jump at 6-hour temporal resolutions. This odd behaviour is most likely related to suboptimal clustering and the consequent distortion of information in the typical profiles. As discussed earlier, clustering coarse temporal resolutions have proven to adversely affect the shape of typical profiles by reducing the range between base and peak loads, eventually distorting the true variability within the consumption patterns. Lastly, heating consumption is not well calibrated when compared to the other metered loads. Although it is difficult to draw a concrete explanation, it is speculated that the heating consumption pattern is strongly affected by the response of the ceiling heating panels, a variable which is not calibrated in this study.

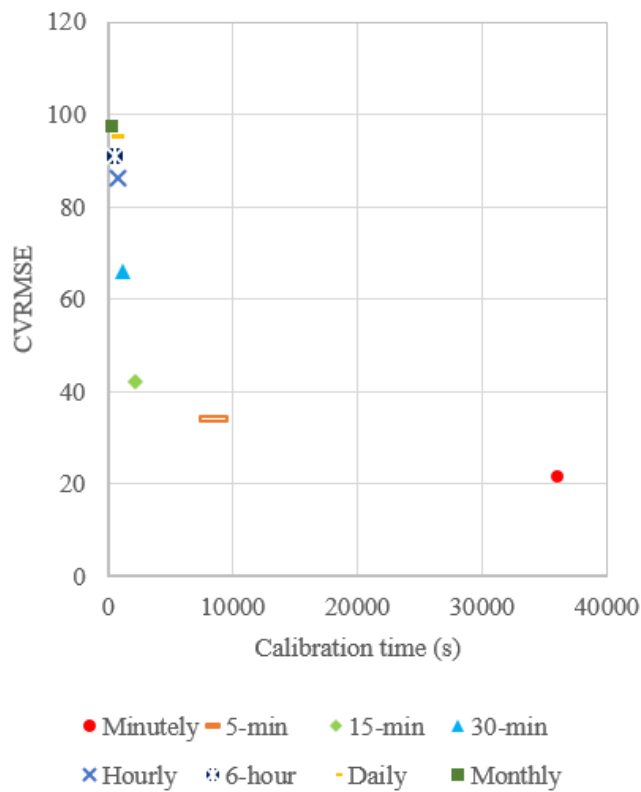

*Figure 3: Calibration accuracy vs computational time for eight temporal resolutions of heating energy consumption.*

To understand how calibration on aggregated data can result in information loss, all calibrated models are run at 1-minute time-steps and contrasted against 1-minute measurements (Table S2). In terms of heating and cooling, not much difference is observed when the calibration temporal resolution is higher than 30-minutes. Electricity calibration seems somewhat more sensitive to the temporal resolution. DHW has the highest sensitivity to the temporal resolutions. Both CVRMSE and NMBE of DHW drastically increase in calibrated models with finer resolutions (Figure S3). Higher accuracy of models calibrated with smaller time-steps comes at the cost of excessive simulation time. Therefore, based on the intended application of calibration, and availability of resources, a trade-off between computational cost and accuracy can be found as shown in Figure 3. For instance, the accuracy of a model calibrated with minutely measurements may seem not much different from that of 15-minute data. However, calibration with 15-minute temporal resolution is roughly 17 times faster than minutely time-steps. The desired solution may differ for electricity and DHW, which are more sensitive to the temporal resolution.

Aside from the accuracy of each model, this study compares the informative priors obtained from each calibration. This allows us to better understand the effects of data aggregation on prior inference. Since DHW energy consumption is solely affected by the DHW peak flow rate and the consumption profile, it is a suitable basis for contrasting different temporal resolutions (Figure 4). It is observed that calibrating with coarser temporal resolutions results in underestimation of the DHW peak flow rate. Coarser temporal resolutions also return smaller variations, which can incorrectly be interpreted as higher confidence. It was previously discussed that coarser temporal resolutions may suffer from information loss, and therefore, the inferred prior should be used with caution. A noticeable gap is observed between temporal resolutions smaller than 6-hour and daily/monthly time-steps. This gap is most likely related to applying the typical SIA profiles to daily and monthly energy models.

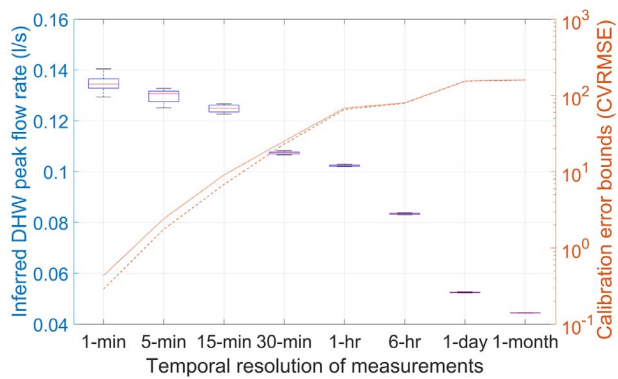

*Figure 4: Inference of prior distributions from calibrated energy models.*

## Conclusion

This study compares the application of different temporal resolutions for calibrating an energy model. Although coarse temporal resolutions tend to show better performances during calibrations, they perform notably worse when contrasted against the ground truth. It is also observed that heating and cooling are less sensitive to the temporal resolutions than electricity and DHW. This is due to the fact that both electricity and DHW consumption are mostly occupant-driven, and therefore, (sub)hourly profiles can greatly affect the reconstruction of the consumption profiles. Finally it is observed that resorting to daily or monthly temporal resolutions for prior inference can be risky, as the true variace within profiles can be greatly distorted when coarse measurements are available.

# Supplementary materials

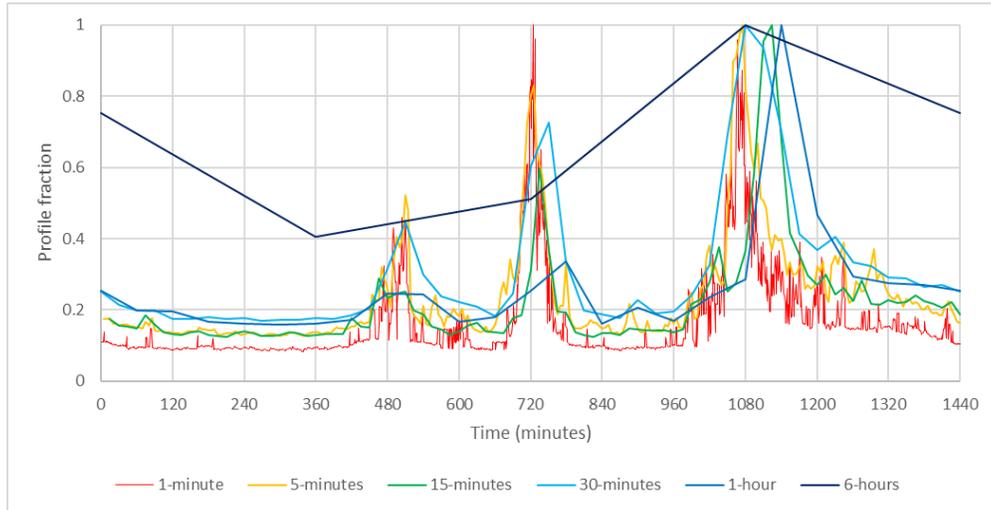

Figure S1: Typical profile of occupancy, lighting and equipment for sample day (Jan 10$^{th}$). Typical days are obtained by clustering the electricity consumption profile.

Table S1: Results of the calibration process. CVRMSE and NMBE are calculated based on measurements that are aggregated to the calibration time-step.

|  | CVRMSE | | | | NMBE | | | |
|---|---|---|---|---|---|---|---|---|
|  | **Heating** | **Cooling** | **Electricity** | **DHW** | **Heating** | **Cooling** | **Electricity** | **DHW** |
| 1-min | 21.58 | 8.87 | 5.11 | 0.29 | 9.15 | 1.67 | 2.92 | 0.06 |
| 5-min | 16.15 | 3.11 | 4.05 | 0.18 | 8.16 | 0.84 | 2.57 | 0.04 |
| 15-min | 9.25 | 1.35 | 1.89 | 0.1 | 5.14 | 0.37 | 1.31 | 0.02 |
| 30-min | 5.48 | 0.72 | 0.99 | 0.06 | 3.18 | 0.19 | 0.74 | 0.01 |
| Hourly | 3.63 | 0.67 | 1.42 | 0.02 | 1.49 | 0.27 | 0.79 | 0.01 |
| 6-hour | 4.53 | 0.6 | 2.92 | 0.01 | 4.46 | 0.51 | 2.9 | 0 |
| Daily | 1.01 | 0.16 | 0.66 | 0 | 0.99 | 0.09 | 0.64 | 0 |
| Monthly | 0.5 | 0.09 | 0.32 | 0 | 0.49 | 0.06 | 0.31 | 0 |

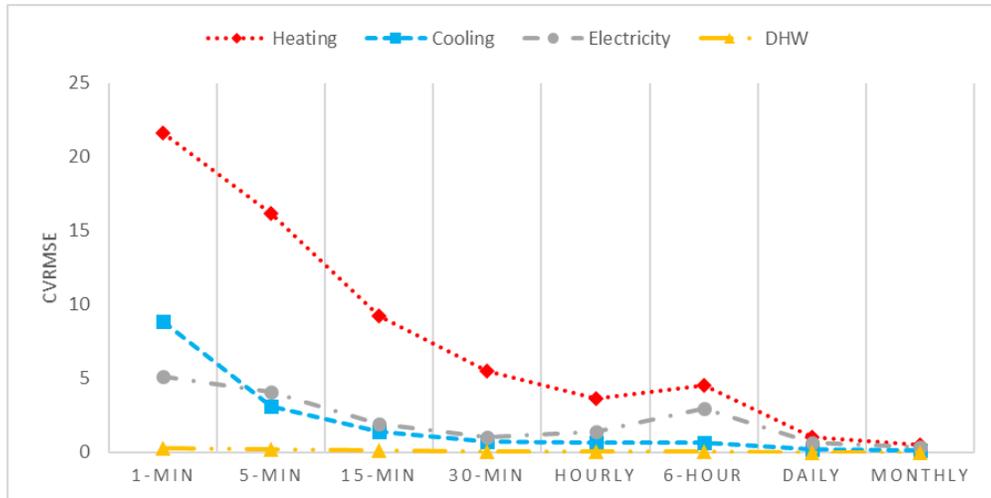

*Figure S2: Calibration error at different temporal resolutions. CVRMSE is calculated between simulation outputs and measurement that are resampled to match the simulation time-step.*

*Table S2: Contrasting calibrated energy models against the ground truth. CVRMSE and NMBE are calculated based on minutely simulation outputs as well as minutely measurements.*

|  | CVRMSE | | | | NMBE | | | |
| ---: | --- | --- | --- | --- | --- | --- | --- | --- |
|  | **Heating** | **Cooling** | **Electricity** | **DHW** | **Heating** | **Cooling** | **Electricity** | **DHW** |
| 1-min | 21.58 | 8.87 | 5.11 | 0.29 | 9.15 | 1.67 | 2.92 | 0.06 |
| 5-min | 33.83 | 20.14 | 8.7 | 1.76 | 13.81 | 7.91 | 4 | 0.08 |
| 15-min | 42.29 | 19.01 | 11.19 | 6.81 | 21.91 | 9.6 | 6.93 | 0.08 |
| 30-min | 65.92 | 37.53 | 24.68 | 23.17 | 22.19 | 13.71 | 12.51 | 3.3 |
| Hourly | 86.13 | 64.21 | 97.63 | 65.55 | 29.98 | 20.97 | 36.54 | 8.76 |
| 6-hour | 91.01 | 73.01 | 129.99 | 79.3 | 30.19 | 29.83 | 39.58 | 30.84 |
| Daily | 95.56 | 81.02 | 136.79 | 153.83 | 33.91 | 34.53 | 41.72 | 136.91 |
| Monthly | 97.47 | 87.37 | 139.18 | 159.64 | 34.44 | 35.29 | 42.05 | 148.75 |

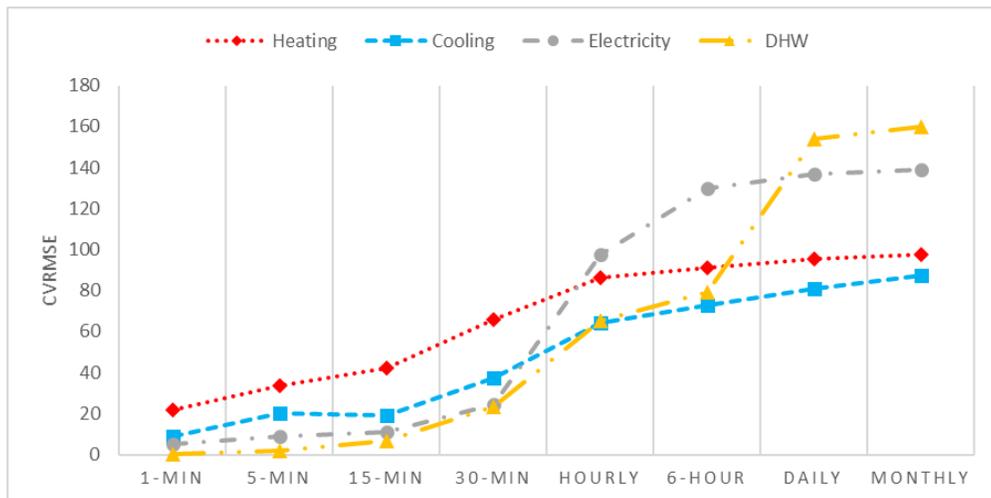

*Figure S3: Temporal resolutions of measurements and the effects on calibrating an energy model. CVRMSE is calculated between simulation outputs resampled to 1-minute temporal resolution and measurements logged at 1-minute intervals.*